\documentclass[aps,preprint,superscriptaddress]{revtex4-2}

\usepackage{graphicx}
\usepackage{amsmath,amssymb}
\usepackage{bm}
\usepackage{hyperref}

\newcommand{\BaNa}{Ba$_{1-x}$Na$_x$Fe$_2$As$_2$}
\newcommand{\BFA}{BaFe$_2$As$_2$}

\DeclareUnicodeCharacter{2212}{-}

\begin{document}
\title{Energy-Resolved Real-Space Imaging of Orbital Nematicity in an Fe-Based Superconductor}

\author{Asato Onishi}
\affiliation{Department of Advanced Materials Science, University of Tokyo, Kashiwa, Chiba 277-8561, Japan}

\author{Zifan Xu}
\affiliation{Department of Advanced Materials Science, University of Tokyo, Kashiwa, Chiba 277-8561, Japan}

\author{C\'edric Bareille}
\thanks{Present address: Hitachi High-Tech Corporation, Hitachinaka, Ibaraki 312-8504, Japan}
\affiliation{Material Innovation Research Center (MIRC), University of Tokyo, Kashiwa, Chiba 277-8561, Japan}

\author{Yoichi Kageyama}
\thanks{Present address: Department of Physics, Kyoto University, Sakyo-ku, 606-8502 Kyoto, Japan}
\affiliation{Department of Advanced Materials Science, University of Tokyo, Kashiwa, Chiba 277-8561, Japan}

\author{Shigeyuki Ishida}
\affiliation{Core Electronics Technology Research Institute, National Institute of Advanced Industrial Science and Technology (AIST), Tsukuba, Ibaraki 305-8568, Japan}

\author{Hiroshi Eisaki}
\affiliation{Core Electronics Technology Research Institute, National Institute of Advanced Industrial Science and Technology (AIST), Tsukuba, Ibaraki 305-8568, Japan}

\author{Kota Ishihara}
\affiliation{Department of Advanced Materials Science, University of Tokyo, Kashiwa, Chiba 277-8561, Japan}

\author{Kenichiro Hashimoto}
\thanks{Present address: Department of Physics, Kyoto University, Sakyo-ku, 606-8502 Kyoto, Japan}
\affiliation{Department of Advanced Materials Science, University of Tokyo, Kashiwa, Chiba 277-8561, Japan}

\author{Toshiyuki Taniuchi}
\affiliation{Material Innovation Research Center (MIRC), University of Tokyo, Kashiwa, Chiba 277-8561, Japan}
\affiliation{Institute for Solid State Physics (ISSP), University of Tokyo, Kashiwa, Chiba 277-8581, Japan}

\author{Takasada Shibauchi}
\affiliation{Department of Advanced Materials Science, University of Tokyo, Kashiwa, Chiba 277-8561, Japan}

\date{\today}

\begin{abstract}
Electronic nematicity in Fe-based superconductors is manifested by spontaneous rotational symmetry breaking and the formation of nematic domains with mutually orthogonal directions of $d_{xz}$/$d_{yz}$ orbital anisotropy.
However, its energy dependence has remained largely unexplored in real space. Using 5.82-eV laser-excited photoemission electron microscopy (laser-PEEM) with an energy-selective slit, we visualize the evolution of linear dichroic (LD) contrast within individual nematic domains of Ba$_{1-x}$Na$_x$Fe$_2$As$_2$ ($x\approx0.08$). We discover a sign reversal of the LD contrast at an energy $\sim0.4$ eV below the Fermi level, directly revealing an inversion of orbital anisotropy inside each domain. This behavior reflects a different energy-dependent redistribution of spectral weight between the $d_{xz}$ and $d_{yz}$ states, highlighting the crucial role of orbital-selective coherence in the nematic phase of Fe-based superconductors.
\end{abstract}

\maketitle

Electronic nematicity, which spontaneously breaks rotational symmetry of the underlying lattice, has been found in a variety of quantum materials~\cite{bohmer_nematicity_2022, shibauchi_exotic_2020, chu_divergent_2012, kasahara_electronic_2012, bohmer_nematic_2014, ishida_novel_2020, mizukami_thermodynamic_2025, eckberg_sixfold_2020, riggs_evidence_2015, sato_thermodynamic_2017, auvray_nematic_2019, murayama_diagonal_2019, ishida_divergent_2020, asaba_evidence_2024, rubio-verdu_moire_2022}.
In Fe-based superconductors (FeSCs), various experiments have revealed significant nematic instability and anisotropic transport properties, accompanied by tetragonal to orthorhombic structural distortion.
Angle-resolved photoemission spectroscopy (ARPES) measurements resolved a band splitting of $d_{xz}$ and $d_{yz}$ orbitals, which are degenerated in the normal tetragonal phase, revealing a ferro-orbital nature of the nematic phase in FeSCs~\cite{nakayama_reconstruction_2014, shimojima_lifting_2014, zhang_observation_2015, watson_emergence_2015, yi_nematic_2019, pfau_momentum_2019}.

One of the key ingredients of the characteristic features of FeSCs is the multi-orbital electronic structure, which leads to intriguing properties such as orbital selectivity~\cite{fanfarillo_nematicity_2017, yu_orbital_2018, PhysRevX.6.021032, de_medici_hunds_2011, yin_kinetic_2011, fanfarillo_nematic_2023}. This is related to the Hund's metal physics and the effect of electronic correlations on the nematic state can lead to $d_{xz}/d_{yz}$ orbital-dependent coherence and renormalization. It is therefore important to study the energy dependence of orbital nematicity in the nematic phase of FeSCs, but this is less explored due to the experimental limitations. 

Because energetically degenerated twin domains (nematic domains) are formed in the nematic phase, detwinning of these domains using an uniaxial strain is needed in ARPES measurements to reduce mixed contributions from both domains and obtain symmetry-breaking spectra within a single domain. Recent ARPES measurements under such strain indeed found that the $d_{xz}$ orbital lifts upwards to the $d_{yz}$ orbital around $\bar{\Gamma}$ and that a spectral weight transfer from coherent bands near $E_{\rm F}$ to the spectrum in the deeper-energy populated by incoherent bands is larger in the $d_{yz}$ orbital compared to the $d_{xz}$ orbital, indicating a different quasiparticle coherence of the $d_{xz}$ and $d_{yz}$ orbitals~\cite{pfau_quasiparticle_2021, pfau_anisotropic_2021}. While these findings highlight the importance of anisotropic correlations, electronic nematicity is strongly coupled to lattice strain and an external strain can modify electronic structures, and thus the purely intrinsic characters of anisotropy in $d_{xz}$ and $d_{yz}$ remain a subject that has not been directly investigated. Specifically, the previous ARPES report on the anisotropic spectral weight in BaFe$_2$As$_2$, which has nearly concurrent antiferromagnetic (AFM) transition alongside the nematic transition, is made solely for a strain response above the nematic transition in order to prevent obstruction from AFM-originated signals, leaving the characters of the unstrained nematic phase unrevealed~\cite{pfau_anisotropic_2021}.

\begin{figure*}[t]
	\includegraphics[clip,width=\linewidth,page=1]{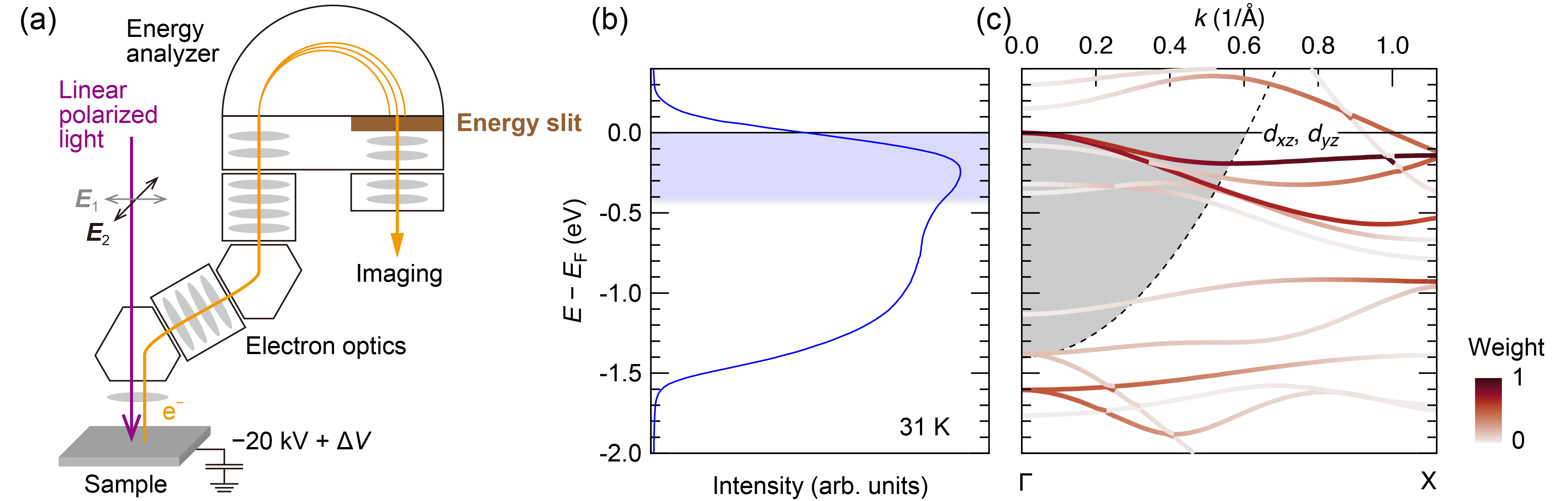}
    \caption{
    Energy-dependent laser-PEEM measurements.
    (a) Schematic of energy-dependent laser-PEEM measurements. The incident light is normal to the sample surface.
    (b) Energy dependence of the laser-PEEM intensity $I_1 + I_2$ obtained at 31\,K for the {\BaNa} sample ($x\approx 0.08$).
    The energy resolution is determined to be $\Delta E \sim 0.2\,$eV by fitting the Fermi step by a Fermi-Dirac distribution convoluted with a Gaussian of full width at half height $\Delta E$, modelizing the response function of the instrument.
    The energy region of $E - E_{\rm F} > -0.4\,$eV is highlighted with blue shade.
    (c) The observable energy-momentum region in PEEM measurements using a $5.82\,$eV laser (gray shaded area), superimposed by the electronic bands of BaFe$_2$As$_2$ obtained from density functional theory (DFT), with a color scale representing the projected weights of $d_{xz}$ and $d_{yz}$ orbitals.
    The DFT calculation is provided by Materials Project database.\cite{jain_commentary_2013}
    The bands are scaled by the renormalization factor of 2 to match previously reported ARPES measurements.\cite{razzoli_tuning_2015}
    }\label{fig_PEEM_schematics}
\end{figure*}

Another approach is to resolve the real-space variations of electronic structure without external strain, which leads to the visualization of nematic domains. A pioneering work on scanning photoelectron spectromicroscopy by Mizokawa {\it et al.} using submicron-focused photon revealed domain structures in an energy range of $-0.5$\,eV to Fermi energy in an Fe-chalcogenide Fe$_{1+\delta}$Te~\cite{mizokawa_mesoscopic_2016}. However, the detailed energy dependence of $d_{xz}$/$d_{yz}$ orbital anisotropy is not resolved in their study.
Although a couple of nano-ARPES measurements followed this work to resolve band and orbital characters within different nematic domains using focused photons, the investigation was limited in the energy range of several hundreds of meV from $E_{\rm F}$.\cite{watson_probing_2019, rhodes_revealing_2020}
More recently, laser-PEEM~\cite{taniuchi_ultrahigh-spatial-resolution_2015} enables high-resolution observation of twinned nematic domains in unstrained samples~\cite{shimojima_discovery_2021, kageyama_coherence_2024}.
PEEM maps the intensity of excited photoelectrons in real space, which depends on the symmetry of the initial electronic state and the incident light.
Linear dichroic (LD) PEEM measurements using two linearly polarized lights along the tetragonal $[110]$ and $[1\bar{1}0]$ directions detect imbalance in density of states between $d_{xz}$ and $d_{yz}$ orbitals: $n_{xz}$ and $n_{yz}$.
As the characteristic directions of $d_{xz}$/$d_{yz}$ orbitals are rotated by $90^\circ$ between the twinned domains, LD imaging visualizes mesoscopic stripe or wave-like nematic domains.
While previous laser-PEEM measurements selectively detected electronic states near $E_{\rm F}$ using a laser of $h\nu=4.66\,$eV, which is close to the typical work function $\phi \sim 4.5\,$eV of FeSCs,~\cite{jung_effect_2021, jiao_significantly_2024} nematic characters of deeper-energy bands have not been accessed due to the limitation of the observable energy region $|E - E_{\rm F}| < h\nu - \phi$, which is a consequence of energy conservation.

In this study, we employed laser-PEEM measurements using a continuous wave $h\nu=5.82\,$eV ($\lambda=213\,$nm) laser to investigate nematic domains in a single crystal of {\BaNa} ($x\approx 0.08$), which is slightly hole-doped, with the nematic transition at $T_{\rm s}\approx 135\,$K.
Because the laser energy is $1.16\,$eV higher than the previous laser-PEEM measurements, photoelectrons from the deeper-energy region at $E - E_{\rm F} \lesssim -0.5\,$eV near $\bar{\Gamma}$, which are reported to be predominantly incoherent in previous ARPES studies,~\cite{evtushinsky_high-energy_2017, watson_formation_2017, pfau_quasiparticle_2021} were resolved by performing energy-dependent acquisition.
While LD images detected coexisting stripe-like nematic domains, the energy dependence of LD signals revealed that the sign of orbital anisotropy is reversed at above and below $E - E_{\rm F} \approx -0.4\,$eV.
These results directly prove that the orbital-dependent spectral weight transfer is present within the coexisting twin domains, without external strain-induced anisotropy.

Figure\,\ref{fig_PEEM_schematics}(a) shows the schematic of energy-dependent laser-PEEM measurement setups.
The incident light is a continuous-wave ultraviolet-laser of $h\nu=5.82\,$eV and enters normal to the sample's $ab$-surface.
By using two linearly polarized lights with polarization vectors $\bm{E}_1$ and $\bm{E}_2$, which are perpendicular to each other, we measure spatially-resolved polarization-dependent PEEM intensity $I_1$ and $I_2$, respectively.
As illustrated in Fig.\,\ref{fig_PEEM_schematics}(a), an energy slit inserted at the end of the energy analyzer limits the window range of detectable energy.
By sweeping the bias voltage $\Delta V$ to the sample, which is an added voltage to the base acceleration voltage of $-20\,$kV, we control the energy range of photoelectrons that pass the window and thus obtain the energy dependence of $I_1$ and $I_2$.

\begin{figure*}[t]
	\includegraphics[clip,width=\linewidth,page=1]{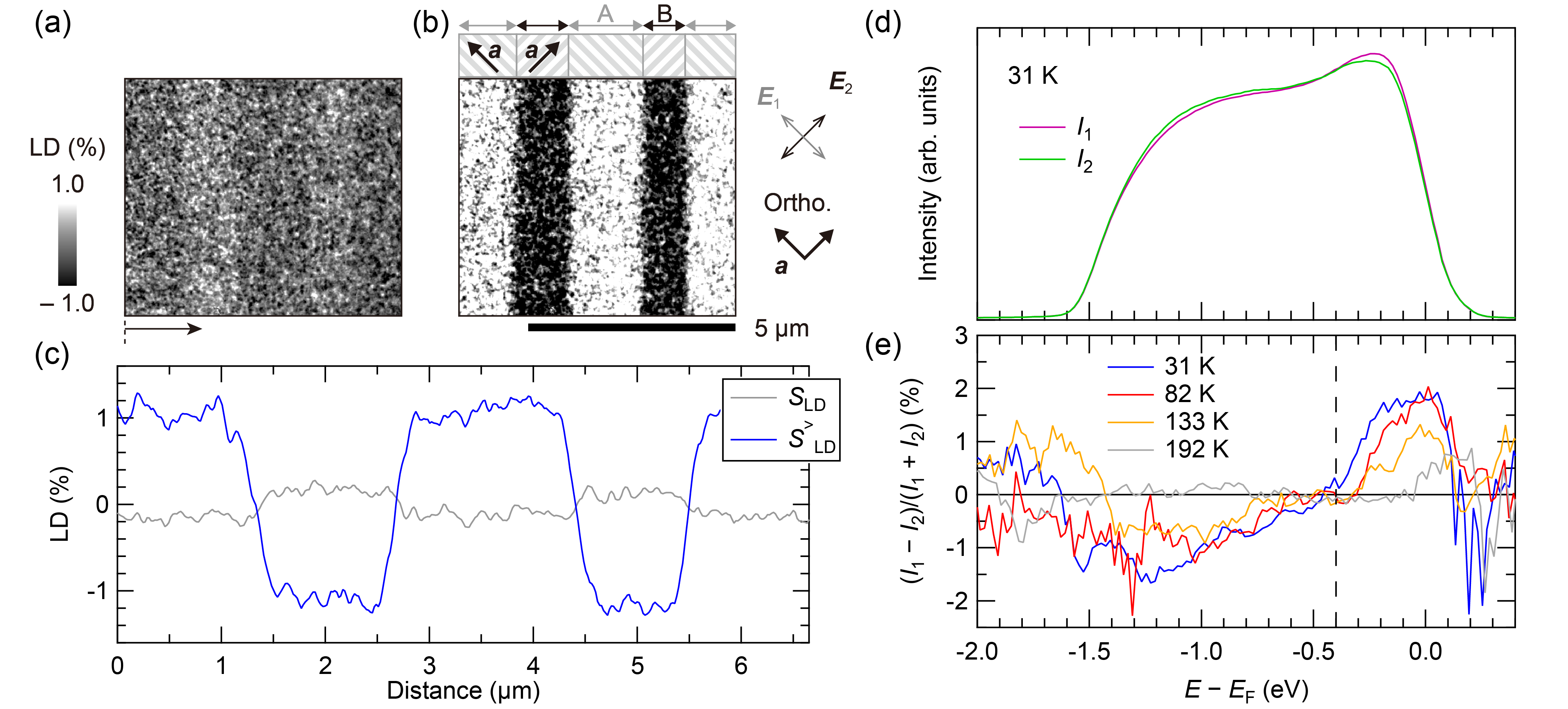}
    \caption{
    Energy-dependent LD measurements.
    (a), (b) Spatial mapping of LD intensity $S_{\rm LD}$ (a) and $S^{>}_{\rm LD}$ (b) at 31\,K, which are obtained from accumulating $I_i$ ($i=1, 2$) for the entire observable energy region and for $E-E_{\rm F} > -0.4\,$eV, respectively.
    The black bar indicates $5\mu$m.
    The schematics above the image in (b) illustrates the structural domains corresponding to the nematic domains where orthorhombic $a$-axis is rotated by $90^\circ$ between different domains.
    (c) Spatial dependence of LD signals obtained by averaging the data shown in (a) and (b) along the stripe direction.
    The black arrow shown below (a) indicates the direction along which the distance is measured.
    (d) Energy dependence of $I_1$ and $I_2$ in the area A, which is indicated in (b).
    (e) Energy dependence of $(I_1 - I_2)/(I_1 + I_2)$ in the area A measured at different temperatures.
    }\label{fig_energy_dependent_LD}
\end{figure*}

Figure\,\ref{fig_PEEM_schematics}(b) shows the energy dependence of PEEM intensity ($I_1 + I_2$) obtained at 31\,K for the {\BaNa} ($x\approx 0.08$) sample.
We first note that the sudden drop of the intensity at $E-E_{\rm F} = -1.4$--$-1.3\,$eV is due to the energy conservation law $E-E_\mathrm{F} + h\nu = E_\mathrm{kin} + \phi$, 
where $E_\mathrm{kin} > 0$ is the kinetic energy of the photoelectron and $\phi$ is the surface work function.
The deepest value of observable $E - E_{\rm F} = \phi - h\nu \sim -1.4$--$-1.3\,$eV is consistent with the typical range of work function in FeSCs ($\phi \sim 4.5\,$eV),~\cite{jung_effect_2021, jiao_significantly_2024} considering the light energy $h\nu=5.82\,$eV.
Importantly, the energy conservation also leads $k_{\parallel}[1/\text{\r{A}}]=0.5123\sqrt{E_\mathrm{kin}\text{[eV]}}\sin{\theta}$, where $\theta$ is the angle (to the sample surface's normal) of photoemission, which limits the observable $k_{\parallel}$ (parallel to the $ab$-plane) to be near the $\bar{\Gamma}$.
Figure\,\ref{fig_PEEM_schematics}(c) illustrates the corresponding observable energy-momentum region, superimposed by the electronic bands calculated for BaFe$_2$As$_2$ displayed with renormalization factor of $\sim2$.\cite{razzoli_tuning_2015}

Figure\,\ref{fig_PEEM_schematics}(b) reveals a peak structure in the energy region of $E-E_{\rm F} \gtrsim -0.5\,$eV, besides a broad hump in the deeper-energy region, which is affected by the energy conservation cutoff.
The bottom of the peak structure is at $E-E_{\rm F} \sim -0.5\,$eV and coincides with the bottom of the coherent electronic bands reported in previous ARPES measurements on a variety of FeSCs, including FeSe and BeFe$_2$As$_2$,~\cite{evtushinsky_high-energy_2017, watson_formation_2017} which revealed that well-defined band dispersions are present only above $\sim -0.5\,$eV (see Fig.\,\ref{fig_PEEM_schematics}(c)) and the spectra get broad and smeared out in the deeper-energy region ($E- E_{\rm F} \lesssim -0.5\,$eV).

We aligned $\bm{E}_1$ to be parallel to the orthorhombic $a$-axis (or equivalently, $b$-axis of the opposite domains) direction so that LD intensity $I_{\rm LD}=(I_1 - I_2)/(I_1 + I_2)$ measures the nematic anisotropy.
Figure\,\ref{fig_energy_dependent_LD}(a) shows an LD image $S_{\rm LD}=(S_1 - S_2)/(S_1 + S_2)$ obtained from summarizing the energy dependence in the entire measurable energy region: $S_i = \sum_E I_i (E)$ ($i=1, 2$).
In parallel, we obtained a similar LD image $S^{>}_{\rm LD}=(S^{>}_1 - S^{>}_2)/(S^{>}_1 + S^{>}_2)$, which is shown in Fig.\,\ref{fig_energy_dependent_LD}(b)), selectively from the spectrum for $E-E_{\rm F} > -0.4\,$eV.
Here we defined $S^{>}_i = \sum_{E-E_{\rm F} > -0.4\,\rm eV} I_i (E)$.
We will see in the later that the threshold of $E-E_{\rm F} > -0.4\,\rm eV$ is valid.
While stripe-like nematic domains are clearly visualized as the contrast of LD intensity $S^{>}_{\rm LD}$ in Fig.\,\ref{fig_energy_dependent_LD}(b), similarly to the previous laser-PEEM studies~\cite{shimojima_discovery_2021, kageyama_coherence_2024}, the contrast is weaker in $S_{\rm LD}$ (Fig.\,\ref{fig_energy_dependent_LD}(b)).
The difference between the two images is also visualized in the spatial dependence of LD averaged along the stripe direction shown in Fig.\,\ref{fig_energy_dependent_LD}(c).
The reduced contrast in $S_{\rm LD}$ indicates that the anisotropy in the energy region close to $E_{\rm F}$ is compensated by the contribution from the deeper-energy region, suggesting an energy-dependent sign change of anisotropy.

To investigate the suggested feature, we grouped the observed stripe-like domains into area A and B, as indicated in Fig.\,\ref{fig_energy_dependent_LD}(b).
Figure\,\ref{fig_energy_dependent_LD}(d) shows the energy dependence of $I_1$ and $I_2$ averaged only for the area A.
While $I_1 > I_2$ in the peak above $E-E_{\rm F} \approx -0.4\,$eV, the dominance is reversed below $E-E_{\rm F} \approx -0.4\,$eV.
The energy dependence of $(I_1-I_2)/(I_1+I_2)$ is obtained for area A and shown in Fig.\,\ref{fig_energy_dependent_LD}(e) for different temperatures, including below and above the transition temperature $T_{\rm s}\approx 135\,$K.
Figure\,\ref{fig_energy_dependent_LD}(e) reveals an energy-dependent sign reversal of LD anisotropy in the nematic phase.
The temperature evolution of the energy dependence features that the sign reversal occurs at $E-E_{\rm F} \approx -0.4\,$eV independently of temperatures, which matches the bottom of the peak structure highlighted in Fig.\,\ref{fig_PEEM_schematics}(b).

As the previous ARPES measurements revealed nematic band-splitting that lifts the $d_{xz}$ band upward near $E_{\rm F}$ and $\Gamma$,~\cite{nakayama_reconstruction_2014, shimojima_lifting_2014, zhang_observation_2015, watson_emergence_2015, yi_nematic_2019, pfau_momentum_2019} we expect $n_{xz} > n_{yz}$ for the coherent band dispersion close to $E_{\rm F}$ ($E-E_{\rm F} > -0.4\,$eV).
Because the linearly polarized light along the orthorhombic $a$ axis is assumed to dominantly detect $d_{xz}$ bands, the positive (negative) $S^>_{\rm LD}$ in the area A (B) is explained by that the direction of the $a$ axis is parallel to the light polarization $\bm{E}_1$ ($\bm{E}_2$) (Fig.\,\ref{fig_energy_dependent_LD}(b)).
Contrastingly, previous ARPES measurements fail to resolve well-defined nematic splitting in the deeper-energy region where the photoemission spectra get smeared out~\cite{evtushinsky_high-energy_2017, watson_formation_2017, pfau_quasiparticle_2021}.
Interestingly however, ARPES measurements using strained FeSCs~\cite{pfau_quasiparticle_2021, pfau_anisotropic_2021} suggest that the photoemission spectral weight of the $d_{yz}$-like component in the deeper-energy region dominates the $d_{xz}$-like one, while the dominance is opposite above $E-E_{\rm F} \approx -0.4\,$eV.
These measurements suggest that the coherence of the well-defined $d_{xz}$ bands close to $E_{\rm F}$ are stronger than the $d_{yz}$ bands in the nematic state, because photoemission spectral weights are assumed to be redistributed from coherent bands to deeper-energy incoherent components as a consequence of the sum rule, depending on the strength of correlation.
The sign reversal observed in Fig.\,\ref{fig_energy_dependent_LD}(e) match well with such an anisotropic coherence between $d_{xz}$ and $d_{yz}$ orbitals, including the energy position of sign change around $-0.4\,$eV.
We note that our work is the first observation for unstrained coexisting nematic domains, which suggest that the anisotropic spectral weight transfer from the coherent bands near $E_{\rm F}$ to the predominantly incoherent dispersions is resolved in real-space.
We also note that the nearly compensated LD shown in Fig.\,\ref{fig_PEEM_schematics}(a) is also consistent with the spectral-weight transfer.

\begin{figure}
	\includegraphics[clip,width=0.5\linewidth,page=1]{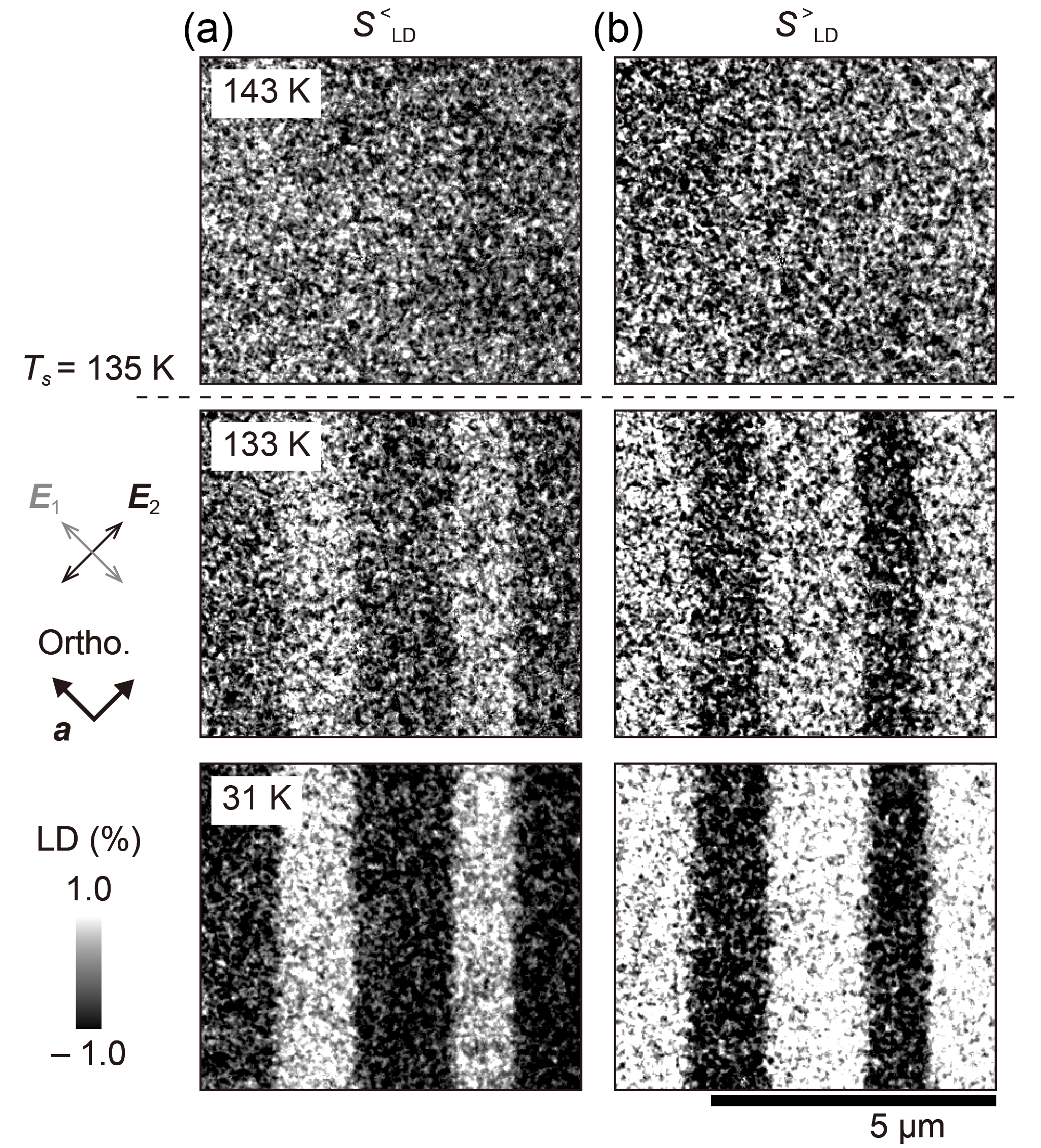}
    \caption{
    LD images at different temperatures obtained from the energy dependence in (a) incoherent ($E-E_{\rm F} < -0.4\,$eV) and (b) coherent ($E-E_{\rm F} > -0.4\,$eV) energy regions.
    }\label{fig_LD213}
\end{figure}

\begin{figure}
	\includegraphics[clip,width=0.5\linewidth,page=1]{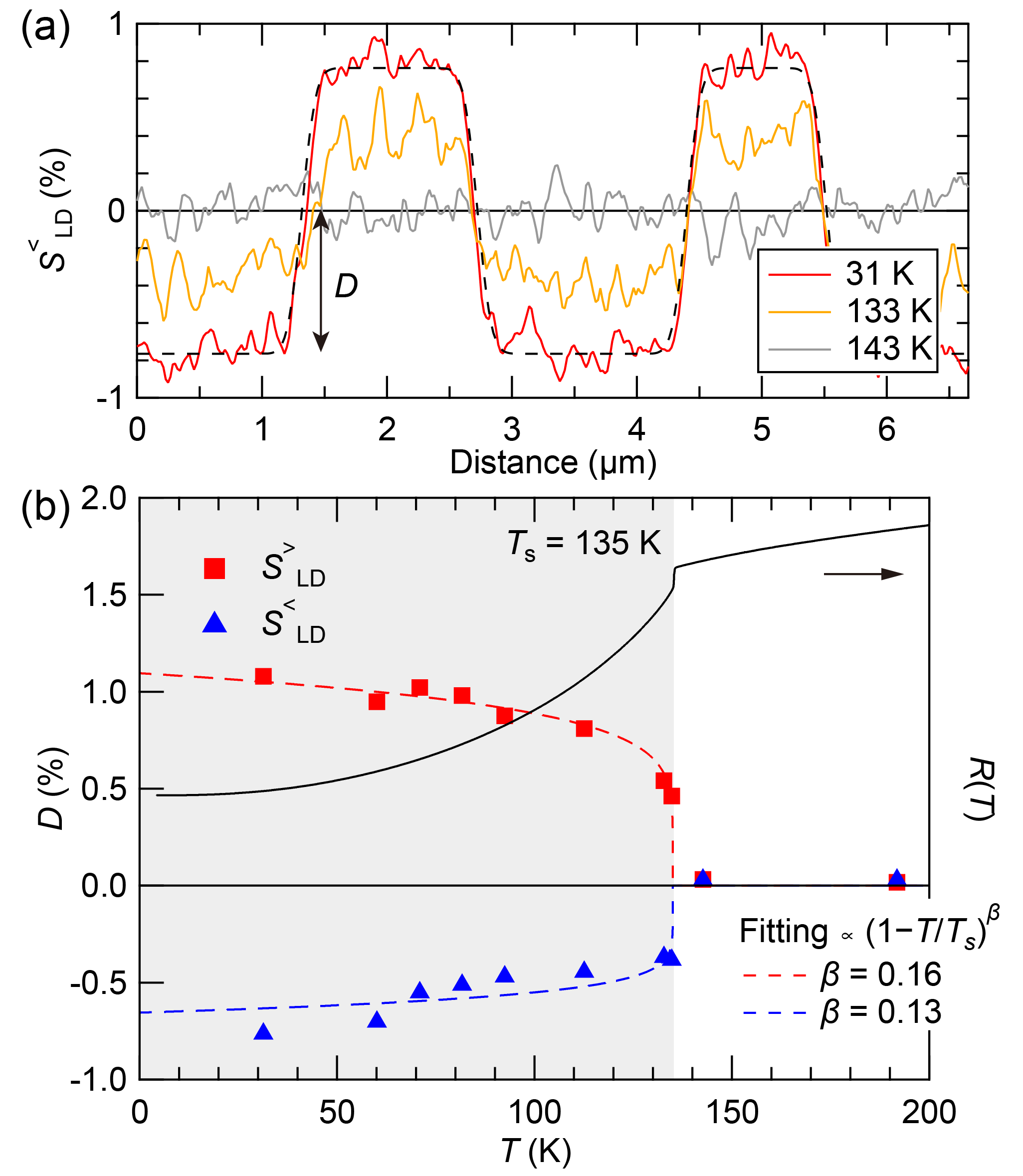}
    \caption{
    Temperature dependence of LD signals.
    (a) Spatial dependence of $S^{<}_{\rm LD}$ at different temperatures, obtained by averaging the spatial distribution along the stripe direction.
    The gray dashed line indicates the fitting.
    (b) The temperature dependence of the fitting parameter $D$ indicated in (a), which measures the contrast of the intensity between nematic domains.
    The red and blue dashed lines indicate fitting functions $\propto (1-T/T_{\rm s})^{\beta}$ for the temperature evolution of $S^{>}_{\rm LD}$ and $S^{<}_{\rm LD}$, respectively.
    The temperature dependence of resistivity is shown on the right axis.
    }\label{fig_stvLD213_Tdep}
\end{figure}

As we have seen that the spectral weight is transferred from the coherent energy region above $E-E_{\rm F}\approx -0.4\,$eV to the predominantly incoherent spectrum below $-0.4\,$eV, let us visualize real-space mapping of LDs for the components above and below $E-E_{\rm F}\approx -0.4\,$eV.
As a counterpart of the $S^{\rm >}_{\rm LD}$, which is introduced above, we define an LD in the predominantly incoherent component below $E-E_{\rm F}\approx -0.4\,$eV: $S^{<}_{\rm LD} = (S^{<}_1 - S^{<}_2)/(S^{<}_1 + S^{<}_2)$, where $S^{<}_i = \sum_{E-E_{\rm F} < -0.4\,\rm eV} I_i (E)$ ($i=1, 2$).
Figure\,\ref{fig_LD213}(a) and (b) show images of $S^{<}_{\rm LD}$ and $S^{>}_{\rm LD}$, respectively.
The images demonstrate that nematic anisotropy emerges below the transition temperature $T_{\rm s}=135\,$K, and that the same domain structure appears with opposite signs in $S^{<}_{\rm LD}$ and $S^{>}_{\rm LD}$ images, while the domain pattern is preserved upon temperature evolution.
Figure\,\ref{fig_stvLD213_Tdep}(a) shows the temperature evolution of $S^{<}_{\rm LD}$ averaged along the stripe direction.
We fitted the square-wave-like spatial dependence of $S^{<}_{\rm LD}$ using a fitting parameter $D$ indicated in Fig.\,\ref{fig_stvLD213_Tdep}(a), which measures the half size of the LD contrast across the nematic domains.
The fitted function is shown as the dashed line in Fig.\,\ref{fig_stvLD213_Tdep}(a).
Figure\,\ref{fig_stvLD213_Tdep}(b) shows the temperature dependence of the LD contrast obtained from $S^{>}_{\rm LD}$ and $S^{<}_{\rm LD}$.
The LD signals $S^{>}_{\rm LD}$ and $S^{<}_{\rm LD}$ evolve with different signs and on a similar scale, further verifying the anisotropic spectral-weight transfer from the coherent bands above $E-E_{\rm F}\approx -0.4\,$eV to the predominantly incoherent spectrum in the deeper-energy.
The onset temperature of $S^{>}_{\rm LD}$ and $S^{<}_{\rm LD}$ match well with the transition temperature $T_{\rm s}=135\,$K evident in the temperature dependence of resistivity (Fig.\,\ref{fig_stvLD213_Tdep}(b)).
Power law fitting $\propto (1-T/T_{\rm s})^{\beta}$ for the temperature dependence of LD signals gives $\beta=0.16(1)$ and $\beta=0.13(4)$ for $S^{>}_{\rm LD}$ and $S^{<}_{\rm LD}$, respectively, which are close to the theoretical expectation $\beta=0.125$ for the magnetic order parameter and the reported value $\beta\approx 0.103$ by a neutron diffraction study in {\BFA}~\cite{wilson_neutron_2009}.
These results indicate that the energy-dependent redistribution of spectral weight in $d_{xz}$ and $d_{yz}$ states sets in just below the nematic transition. 

In summary, we have demonstrated the real-space visualization of the orbital anisotropy reversal at $\sim 0.4$\,eV below the Fermi energy in the nematic domains of an FeSC. This strongly suggests the $d_{xz}/d_{yz}$ orbital dependent coherence and energy transfer below the nematic transition. Such orbital selectivity in the nematic phase of FeSCs is a crucial element in discussing anisotropic quasiparticle interference and the superconducting gap, serving as a key to understanding the microscopic mechanisms of nematicity and superconductivity.

\begin{acknowledgments}
We thank the fruitful discussions with Hiroshi Kontani and Youichi Yamakawa.
This work was supported by a Grant-in-Aid for Transformative Research Areas (A) ``Correlation Design Science'' (No.\ JP25H01248), by Grants-in-Aid for Scientific Research (KAKENHI) (Nos.\ JP22H00105 and JP25H00838) from Japan Society for the Promotion of Science, and by JST SPRING (Grant No.\ JPMJSP2108).
\end{acknowledgments}

\bibliography{BaNa122_JPSJ2025_refs}

\end{document}